\documentclass[12pt,preprint]{aastex}
\usepackage{natbib,color}
\newcommand{\myemail}{guoyang@nju.edu.cn}

\begin{document}
\title{MHD Seismology of a Coronal Loop System by the First Two Modes of Standing Kink Waves}
\author{Y. Guo$^{1}$, R. Erd\'elyi$^2$, A. K. Srivastava$^3$, Q. Hao$^{1}$, X. Cheng$^{1}$, P. F. Chen$^{1}$, M. D. Ding$^{1}$, B.~N. Dwivedi$^{3}$}

\affil{$^1$ School of Astronomy and Space Science \& Key Laboratory of Modern Astronomy and Astrophysics in Ministry of Education, Nanjing University, Nanjing
210046, China} \email{\myemail}

\affil{$^2$ Solar Physics and Space Plasma Research Center (SP$^2$RC), School of Mathematics and Statistics, University of Sheffield, Sheffield, S3 7RH, UK}

\affil{$^3$ Department of Physics, Indian Institute of Technology (Banaras Hindu University), Varanasi-221005, India}

\begin{abstract}
We report the observation of the first two harmonics of the horizontally polarized kink waves excited in a coronal loop system lying at south-east of AR 11719 on 2013 April 11. The detected periods of the fundamental mode ($P_1$), its first overtone ($P_2$) in the northern half, and that in the southern one are $530.2 \pm 13.3$, $300.4 \pm 27.7$, and $334.7 \pm 22.1$ s, respectively. The periods of the first overtone in the two halves are the same considering uncertainties in the measurement. We estimate the average electron density, temperature, and length of the loop system as $(5.1 \pm 0.8) \times 10^8$ cm$^{-3}$, $0.65 \pm 0.06$ MK, and $203.8 \pm 13.8$ Mm, respectively. As a zeroth order estimation, the magnetic field strength, $B = 8.2 \pm 1.0$ G, derived by the coronal seismology using the fundamental kink mode matches with that derived by a potential field model. The extrapolation model also shows the asymmetric and nonuniform distribution of the magnetic field along the coronal loop. Using the amplitude profile distributions of both the fundamental mode and its first overtone, we observe that the antinode positions of both the fundamental mode and its first overtone shift towards the weak field region along the coronal loop. The results indicate that the density stratification and the temperature difference effects are larger than the magnetic field variation effect on the period ratio. On the other hand, the magnetic field variation has a greater effect on the eigenfunction of the first overtone than the density stratification does for this case.
\end{abstract}


\section{Introduction}
Solar magneto-seismology adopts magnetohydrodynamic (MHD) wave theories and observations to probe the physical parameters in the corona \citep{1983Edwin,
1984Roberts,2005Nakariakov,2007Banerjee,2009Andries,2009DeMoortel,2009Ruderman,2013Mathioudakis}. The linear wave properties are described by the dispersion relations of the eigen-modes and the eigenfunction itself in a magnetic cylinder, which is considered as an abstract model of a coronal loop, a filament, or, e.g. a plume. Certain types of MHD waves can be trapped in a magnetic cylinder if the external Alfv\'en speed, $C_\mathrm{Ae}$, is greater than the internal speed, $C_\mathrm{A0}$. Under usual coronal conditions, we have $C_\mathrm{s0} < C_\mathrm{A0} < C_\mathrm{Ae}$, where $C_\mathrm{s0}$ is the sound speed inside the magnetic cylinder. In general, the relationships between $C_\mathrm{s0}$, $C_\mathrm{A0}$, and $C_\mathrm{Ae}$ depend on the plasma parameters, such as the plasma $\beta$ (ratio between the gas pressure and the magnetic pressure), magnetic field distributions, and density distributions. Non-leaky MHD body waves are divided into different categories depending on their characteristic speeds and azimuthal wave numbers, $m$. For instance, in the case of $C_\mathrm{s0} < C_\mathrm{A0} < C_\mathrm{Ae}$, MHD waves with speeds less than $C_\mathrm{s0}$, equal to $C_\mathrm{A0}$, or greater than $C_\mathrm{A0}$, are regarded as slow, torsional Alfv\'en, or fast waves, respectively. For those fast magneto-acoustic waves with $m = 0$ or $1$, they are regarded as fast sausage or kink tubular modes, respectively. In closed coronal magnetic loops, the wavelength could be prescribed by the loop length due to the line-tied conditions in the photosphere. These trapped waves are standing waves in nature and may consist of various modes, such as the fundamental mode and its overtones depending on the longitudinal wave number, $k$. We note that, even in a closed loop, propagating waves could exist if the wavelength is small, or, the propagation time is not long enough to develop a standing mode.

Fast-mode kink oscillations have been observed by the \textit{Transition Region and Coronal Explorer} \citep[\textit{TRACE};][]{1999Aschwanden,1999Schrijver,2004Wang} and other instruments \citep[e.g.,][]{2011Aschwanden,2012White,2013Srivastava}.
Using the properties of kink oscillations, one can infer the coronal magnetic field (\citealt{2001Nakariakov,2008Erdelyi,2008VanDoorsselaere}; and for recent reviews see e.g. \citealt{2009Andries,2009Ruderman}). However, using fundamental frequency alone, one could only estimate the average physical parameters along a coronal loop. Considering overtones, as their frequency ratio to the fundamental frequency usually departs from the canonical integer value for a uniform loop, provides the longitudinal and radial variation information of the physical parameters, such as density, magnetic field strength, and so on \citep{2007Erdelyi,2007Verth,2012Luna-Cardozo,2012Orza,2014Erdelyi}. Overtones of standing kink oscillations have been detected by, e.g. \citet{2004Verwichte}. Due to the relatively short period and small amplitude of kink-mode overtones, only a few more cases have been reported \citep{2007VanDoorsselaere,2007OShea,2008Verth,2011Ballai,2012White,2013Srivastava}. A comprehensive review including the theoretical insight on the coronal seismology by kink-mode overtones can be found in \citet{2009Andries} and \citet{2009Ruderman}.

In this paper, we use high spatio-temporal resolution observations of the Atmospheric Imaging Assembly \citep[AIA;][]{2012Lemen} on board the \textit{Solar Dynamics Observatory} (\textit{SDO}) to analyze the kink oscillations of a coronal loop and identify its first two modes (fundamental mode and its first overtone). The Differential Emission Measure (DEM) analysis provides the temperature and density information of the coronal loop. We also obtain the three-dimensional magnetic structure with a potential field model to find the geometrical parameters and magnetic field strength distributions. As a first and novel step, we verify the consistency of the magnetic field derived by the magnetic field extrapolation and that by the solar magneto-seismology using the fundamental mode of the kink waves. The key point of this paper is to observe the shift of the antinodes for the first overtone relative to that of the fundamental mode when both the density stratification and magnetic field expansion are considered. This enables us to examine physically that which mechanism  is effectively at work on the excited kink oscillations under the framework of spatial MHD seismology. Observations and data analysis are described in Section~\ref{sect2}. Results are presented in Section~\ref{sect3}. Discussion and conclusions are made in Section~\ref{sect4}.

\section{Observations and Data Analysis} \label{sect2}

\subsection{Instrument and Loop Location}

\textit{SDO}/AIA provides full-disk observations of the Sun with high temporal resolution of 12~s and high spatial resolution of $1.5''$ (the pixel sampling is $0.6''$) in seven extreme-ultraviolet (EUV) spectral lines and three UV-visible continua. It covers a large temperature range from 0.06 to 20 MK. \textit{SDO}/AIA consists of four 20 cm telescopes and four cameras that enable quasi-simultaneous observations in all the selected spectral lines. Each detector has $4096 \times 4096$ pixels and the field of view is $41 \times 41$ arcmin$^2$ along the detector axes and $46 \times 46$ arcmin$^2$ along the detector diagonal.

On 2013 April 11, an M6.5 class flare occurred in AR 11719. The flare started at 06:55 UT, peaked at 07:16 UT, and triggered coronal loop oscillations in a quiescent region, located in the south-east direction from the flare epicenter. There were two more active regions, namely, 11721 and 11722, to the south-east of the loop. Figure~\ref{fig1}(a) shows the coronal loop in a 171~\AA \ image recorded by \textit{SDO}/AIA at 06:56 UT. We have checked the observations in all the other wavebands. The coronal loop was most clearly visible in the 171~\AA \ waveband. As shown in Figure~\ref{fig1}(a), the loop spanned over the equator, and the longitude of the loop apex was about $28^\circ$ to the east of the central meridian.

\subsection{Magnetic Field Modelling and 3D Geometry}

\begin{figure}
\begin{center}
\includegraphics[width=1.0\textwidth]{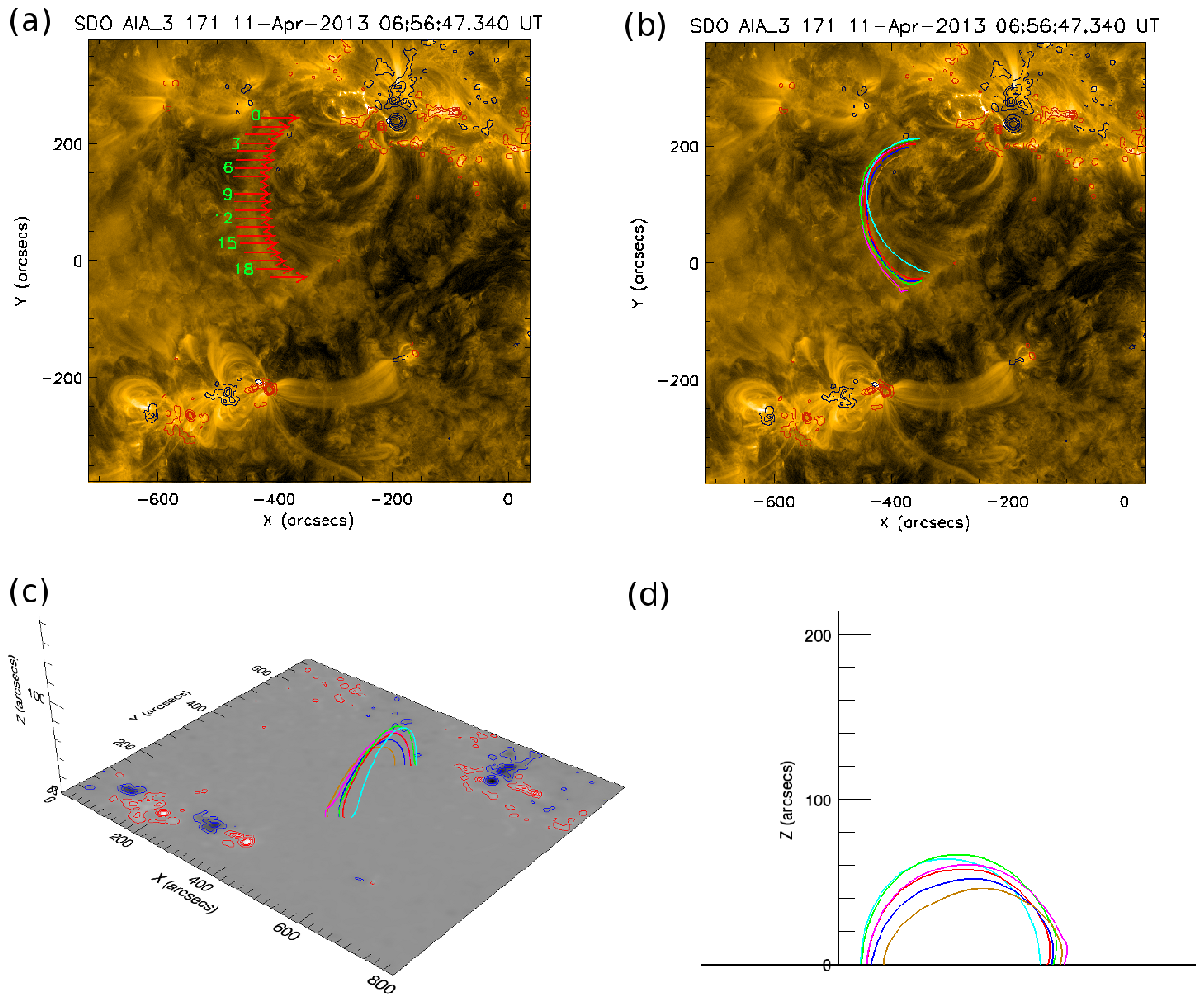}
\caption{
(a) \textit{SDO}/AIA 171~\AA \ image at 06:56 UT on 2013 April 11. Arrows and numbers mark the slices across a coronal loop. The arrow direction indicates the upward direction of the slices. Red and blue contours represent positive and negative magnetic polarities, respectively. (b) \textit{SDO}/AIA 171~\AA \ image overlaid with a potential field on 06:48 UT. Solid lines show the potential field lines modelling the coronal loop. (c) Perspective view of the magnetic loop. Grey-scale image represents the vertical component of the vector magnetic field. (d) An edge-on view of the magnetic loop.
} \label{fig1}
\end{center}
(A movie of the \textit{SDO}/AIA 171~\AA \ observations is available in the online journal.)
\end{figure}

The photospheric vector magnetic field is observed by the Helioseismic and Magnetic Imager \citep[HMI;][]{2012Scherrer,2012Schou} on board \textit{SDO}. We adopt the minimum energy method to resolve the $180^\circ$ ambiguity of the transverse magnetic field vectors \citep{2006Metcalf,2009Leka}. We use the vector magnetic field instead of only the line-of-sight component to correct the projection effect \citep{1990Gary}. A potential magnetic field is computed using the Green's function method \citep{1964Schmidt} as shown in Figure~\ref{fig1}(c). The magnetic field lines modelling the coronal loops observed in 171~\AA \ are selected as follows: first, we integrate several tens of magnetic field lines starting close to both footpoints of the coronal loop. Then, we project the magnetic field back to the \textit{SDO}/AIA viewing angle as shown in Figure~\ref{fig1}(b). Finally, only those magnetic field lines which resemble the 171~\AA \ coronal loop are kept. Figure~\ref{fig1}(b) shows that the coronal loop in the quiescent region is well reconstructed by the potential field.

We have checked the observations of the twin \textit{Solar TErrestrial RElations Observatory} (\textit{STEREO}). None of the spacecrafts observed the region of interest. Therefore, magnetic field modelling seems to be the only method to derive the three-dimensional structure for this case. From Figure~\ref{fig1}(c), we find that the coronal loops are almost vertical to the local horizontal plane. If we estimate the loop length, $L_\mathrm{osc}$, as the average lengths of the sample field lines and the errors as their standard deviation, it is found that $L_\mathrm{osc} = 203.8 \pm 13.8$ Mm. Comparing Figure~\ref{fig1}(b) and \ref{fig1}(d), the height of the loop apex is less than half of the distance between the two footpoints. Therefore, the coronal loop is not a semicircle.

\subsection{Excitation of Loop Oscillation}

The loop oscillation was excited by the global coronal fast magneto-acoustic wave, which was probably generated by the flux rope eruption and associated flare/CME \citep[e.g.,][]{2002Chen,2005Ballai,2012Cheng}. The wave travelled approximately horizontally and drove the oscillation of the coronal loops. Therefore, the oscillation has a horizontal polarization, namely, the loop plane and the oscillation plane are perpendicular to each other.

The propagation speed of the global fast magneto-acoustic driver wave can be estimated following the method proposed by \citet{2011Aschwanden}. We locate the flare position ($x_\mathrm{flare}=-205''$, $y_\mathrm{flare}=+231''$) as the center of the two main polarities as shown in Figure~\ref{fig1}(a). The loop height ($h_\mathrm{apex}=58''$) is estimated as the average of the apex height of the sample field lines. Then, we take the average distances from the flare center to the loop positions as the travel distance, $L_\mathrm{exc}$, which is $229.5 \pm 10.4$ Mm. The sample distances are measured along the great circle at radius $R_\sun + h_\mathrm{apex}$ from $x_\mathrm{flare}=-205''$ and $y_\mathrm{flare}=+231''$ to the 9 crossover points on the slices (number 10 to 18 as shown in Figure~\ref{fig1}(a), where the loop oscillations are clearly observed) with the coronal loop. The error is computed as the standard deviation of the distances. The time delay between the flare start (06:55 UT) and the oscillation start (07:00 UT), $T_\mathrm{exc}$, is about $300 \pm 60$ s, where we adopt an upper limit error of 60 s. The wave speed is thus, $v_\mathrm{exc} = L_\mathrm{exc}/T_\mathrm{exc} = (229,500 \pm 10,400 \ \mathrm{km})/(300 \pm 60 \ \mathrm{s}) = 765 \pm 157 \ \mathrm{km \ s}^{-1}$. This is a typical coronal fast wave speed \citep{2014Liu}.

\begin{figure}
\begin{center}
\includegraphics[width=1.0\textwidth]{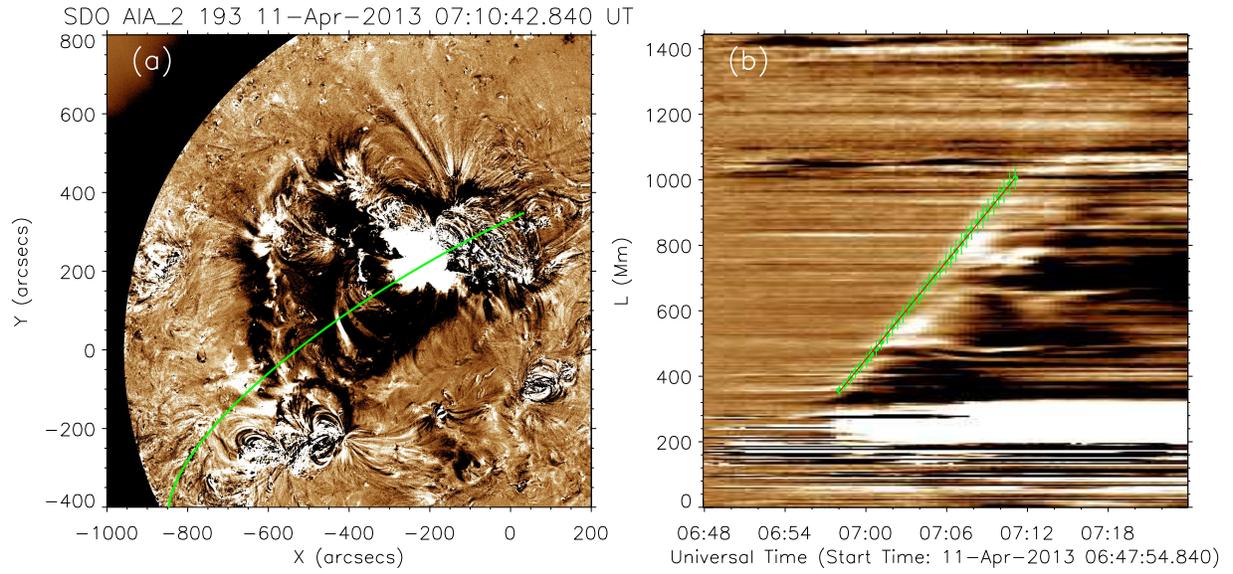}
\caption{
Base difference and time-distance images of 193~\AA . (a) Green solid line indicates an arc-along great circle on the solar surface. It passes the flare center at $x_\mathrm{flare}=-205''$, $y_\mathrm{flare}=+231''$. (b) The slice image is placed such that the southeast direction on the solar surface is upward in the time-distance image. Green dots and vertical lines show the measured position and uncertainties of the propagating EUV wave. Red solid line represents a linear curve fitted to the measurement.
} \label{fig2}
\end{center}
\end{figure}

We estimate the propagation speed of the global fast magneto-acoustic wave with another method to double-check the result derived above. \textit{SDO}/AIA observations show that the EUV waves are most clearly observed in the 193~\AA \ waveband. The observations also show some evidence that there is another propagating wave-like structure after the fast one. The slow wave is recognized as the so-called ``EIT'' wave \citep{2002Chen,2011Chen}. Here, we focus on the fast magneto-acoustic wave and leave the analysis of both waves for another study. Figure~\ref{fig2}(a) shows a base difference image of 193~\AA , which is constructed by subtracting the analyzed image with a fixed reference image observed at 06:47:30 UT on 2013 April 11. \textit{SDO}/AIA adopts a 2 s and a shorter exposure time by turns with the cadence of 12 s to record the quiescent and flare regions quasi-simultaneously. We only use the data with an exposure time of 2 s to highlight the quiescent region, which results in a cadence of 24 s. The slice is selected as the great circle passing the flare center and the coronal loop as shown in Figure~\ref{fig2}(a).

Figure~\ref{fig2}(b) shows the time-distance image in 193~\AA . We quantitatively measure the distance of the fast magneto-acoustic wave by interactively clicking the position of the wave in a uniform time grid with 24 s cadence. We repeat the measurement ten times and compute the distance, $L$, and the uncertainties as the average value and standard deviation, respectively. Next, the measurements are fitted by the first order polynomial,
\begin{eqnarray}
L(t) = v_\mathrm{exc}(t-t_0) + L_0, \label{eqn:v_wave}
\end{eqnarray}
where $t_0$ is 06:57:55 UT, $v_\mathrm{exc} = 825.0 \pm 19.5$ km~s$^{-1}$, and $L_0 = 351.9 \pm 7.2$ Mm. Considering the uncertainties, the fitted velocity of the fast magneto-acoustic wave is consistent with $765 \pm 157 \ \mathrm{km \ s}^{-1}$ derived by the method of \citet{2011Aschwanden}. We will adopt $v_\mathrm{exc} = 825.0 \pm 19.5$ km~s$^{-1}$ to carry out the following analysis. In addition, since the corona is a low plasma-$\beta$ environment, we could take $v_\mathrm{exc}$ as the Alfv\'en speed, $v_\mathrm{Ae}$, external to the coronal loop.

\subsection{Measurement of Loop Oscillation} \label{sect24}

To study the loop oscillations along the loop length, $s$, we select a series of slices across the loop. Our slice selection criterion is that the loop oscillation should have as large amplitude along the slice as possible. Because the loop is nearly in the south-north direction, close to the equator, and the oscillation has a horizontal polarization, the slices are selected as parallel to the east-west direction. For example, in the middle point of the loop, the projection effect shrinks a line in the north-south direction by a factor of about 0.99, which is a negligible effect. We only need to consider the projection effect in the east-west direction, and this consideration greatly simplifies the following analysis. We select twenty slices with equal interval in the directly observed image as shown in Figure~\ref{fig1}(a). The positions will be converted to the length along the loop later.

\begin{figure}
\begin{center}
\includegraphics[width=0.9\textwidth]{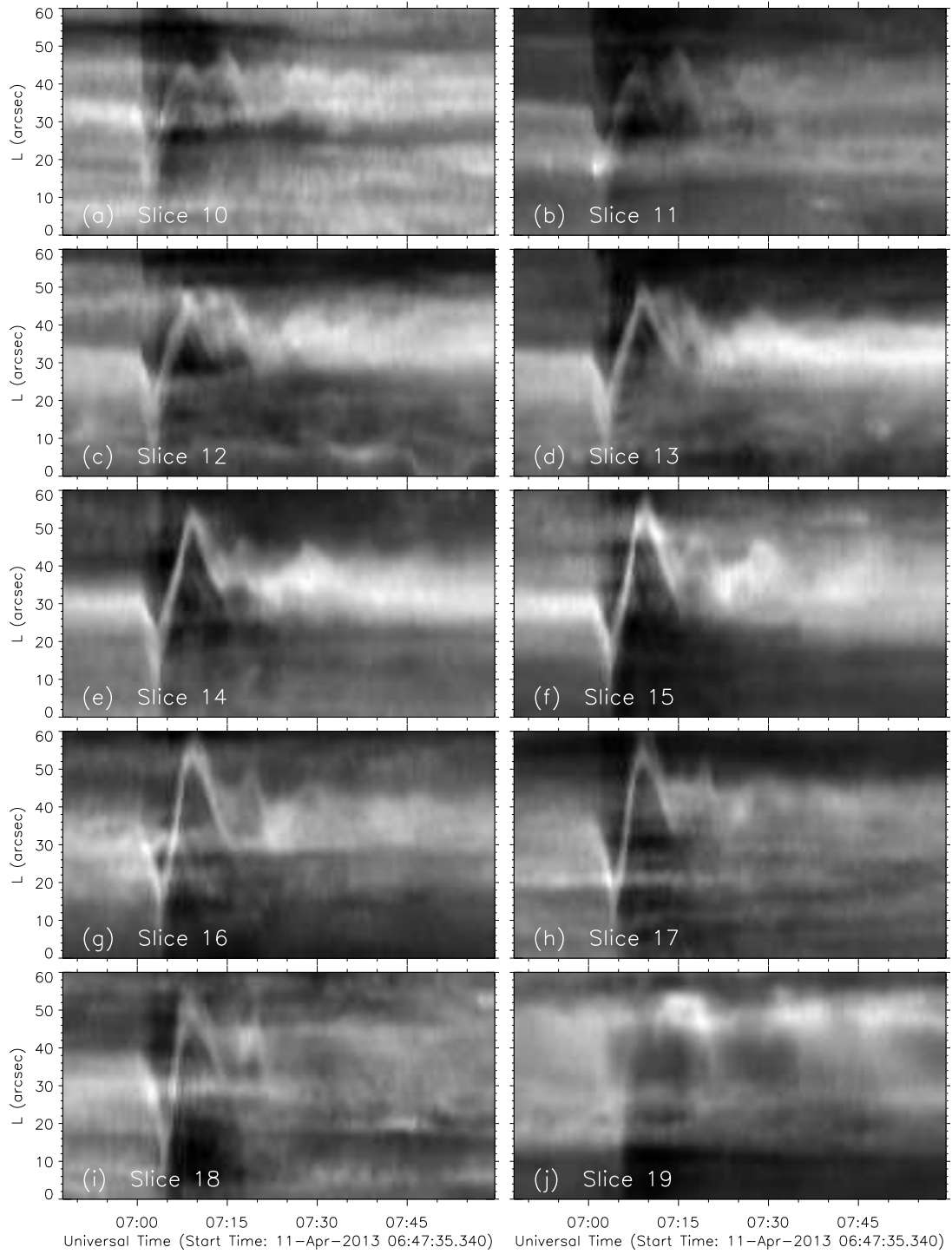}
\caption{
Time-distance image showing the coronal loop oscillations in 171~\AA . The slice positions and numbers are shown in Figure~\ref{fig1}(a).
} \label{fig3}
\end{center}
\end{figure}

We make time-distance images along the twenty slices chosen as shown in Figure~\ref{fig1}(a). The 171~\AA \ images along the selected slices are stacked vertically from left to right according to the time from 06:47 UT to 08:00 UT. Figure~\ref{fig3} shows only ten time-distance images since the oscillations do not appear clearly along the other slices. The time-distance images show several features of the coronal loop oscillations. First, the oscillations in different positions started at almost the same time, 07:00 UT. Secondly, the first shift of the oscillation is away from the flare region. Thirdly, when the loop relaxed to an equilibrium state, it was closer to the flare region than the initial position was. It is due to the decrease of the pressure in the flare region by the flux rope eruption. The coronal loop is pushed to the flare region by the high pressure in the far side. Finally, the oscillations along slices 14--16 have apparent longer periods than that along slices 10--13 and 17--18, which might be due to that overtone signals were detectable along slices 10--13 and 17--18.

\begin{figure}
\begin{center}
\includegraphics[width=0.7\textwidth]{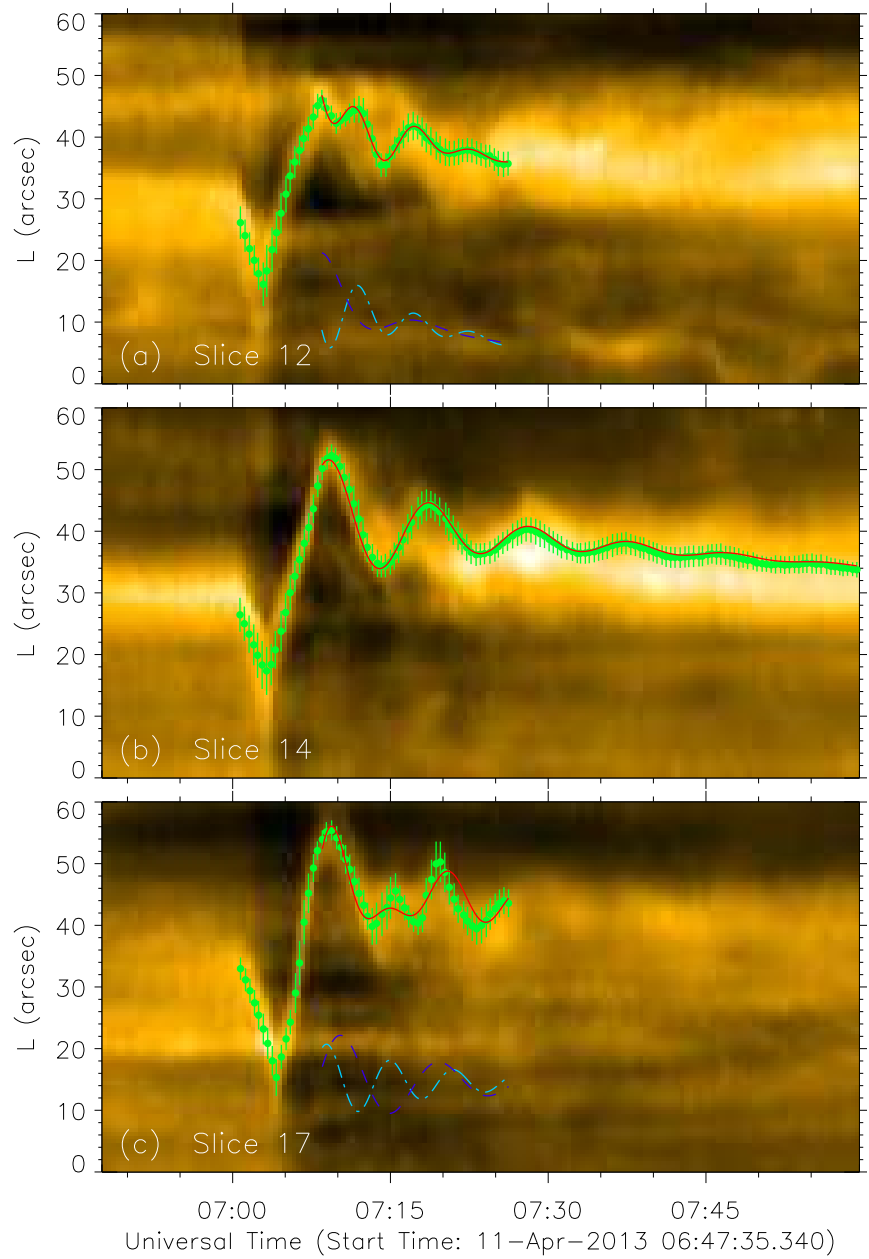}
\caption{
Loop oscillation measurements and fitting results. Filled dots and error bars are loop positions measured from the time-distance image. Solid lines are fitted curves using damping or combined damping cosine profiles. Dash and dash-dotted lines are the two components of the combined damping cosine profiles, which are shifted by an arbitrary distance to a lower position. Panels (a), (b), and (c) are the results for slices 12, 14, and 17, respectively.
} \label{fig4}
\end{center}
\end{figure}

To quantify the oscillation parameters of the coronal loop along different slices, we measure the oscillation profiles by interactively clicking points on the time-distance images. We repeat this process ten times for each slice to estimate the measurement uncertainties, which are computed as the standard deviation of the ten measurements. The measured loop positions and their uncertainties are plotted in Figure~\ref{fig4}. Then, we fit a damping cosine profile to the loop oscillations of slice 14:
\begin{eqnarray}
A(t) = A_{00} + A_{01}(t-t_0) + A_1 \cos [\frac{2\pi}{P_1}(t-t_0) - \phi_{01}]e^{-\frac{t-t_0}{\tau_1}}, \label{eqn:dcosine}
\end{eqnarray}
and a combined damping cosine profile to the loop oscillations of slices 12 and 17:
\begin{eqnarray}
A(t) = A_{00} + A_{01}(t-t_0) + A_1 \cos [\frac{2\pi}{P_1}(t-t_0) - \phi_{01}]e^{-\frac{t-t_0}{\tau_1}} + A_2 \cos [\frac{2\pi}{P_2}(t-t_0) - \phi_{02}]e^{-\frac{t-t_0}{\tau_2}}, \label{eqn:cdcosine}
\end{eqnarray}
where $A$, $t$, $A_{00}$, $A_{01}$, and $t_0$ stand for the measured positions along the slice, time, displacement at the reference time, change rate of the displacement, and reference time, respectively. The other free parameters $A_1$, $P_1$, $\phi_{01}$, $\tau_1$ represent the initial amplitude, period of the oscillation, initial phase, and damping time for the first cosine profile, respectively. The subscript 2 stands for the second one.

To find a reasonable fitting result for slice 12, we prescribe the period for the first damping cosine profile as $P_1 = 520$~s, which equals to the period, $P_1$, for slice 17. Since the loop oscillation along slice 17 does not have a changing displacement, we set $A_{01} = 0$ to minimize the number of free parameters. We have tried the fitting for all the measurements in slices 10--18, where the loop oscillations clearly appear. It shows that slices 12--13, slices 14--16, and slices 17--18 have similar fitting parameters, respectively. Here, only the fitting results for slices 12, 14, and 17 are shown in Figure~\ref{fig4} and listed in Table~\ref{tbl1}.

The fitting results clearly show the existence of spatially resolved fundamental and the first overtone excited in the coronal loop. On the one hand, the fundamental mode exists along all the slices. The fitted oscillation parameters conform with the theory, namely, the amplitude for the fundamental mode ($A_1$) is the largest in a middle point (e.g. slice 14), and smaller towards both ends of the loop (e.g. slices 12 and 17). The period ($P_1$), initial phase ($\phi_{01}$), and damping time ($\tau_1$) are the same at different positions along the loop considering the uncertainties as shown in Table~\ref{tbl1}. On the other hand, the first overtone can also be identified by the fitting results since anti-phase oscillations clearly appear at slices 12 and 17. As listed in Table~\ref{tbl1}, the phase difference between $\phi_{02}$ at slice 12 and 17 is $208.8^\circ \pm 45.3^\circ$, which is consistent with the theoretically predicted value of $180^\circ$. The periods for the first overtone at slices 12 and 17 equal to each other considering the uncertainties. We also find that $P_1/P_2 = 1.7 \pm 0.2$ for slice 12 and $P_1/P_2 = 1.6 \pm 0.2$ for slice 17, both of which are less than 2.

The fitting of the time series does not start from the initiation of the loop oscillation (about 07:00 UT along slices 10--18), but starts at a reference time $t_0$ as listed in Table~\ref{tbl1}. The reference time $t_0$ is about 9 minutes after the initiation of the loop oscillation, and this time range is comparable to the period of the fundamental mode, $P_1$. Therefore, it is enough for the incident and reflected waves to travel back and forth along the coronal loop to build up the standing wave. Besides, the fitting results listed in Table~\ref{tbl1} show that the initial phases, $\phi_{01}$, for the fundamental mode at different places are the same to each other within the measurement uncertainties. The initial phases, $\phi_{02}$, for the first overtone at the two parts of the coronal loop are out of phase. These findings also support that the standing waves have been built up after the reference time $t_0$.

\subsection{Loop Density and Temperature}

\begin{figure}
\begin{center}
\includegraphics[width=1.0\textwidth]{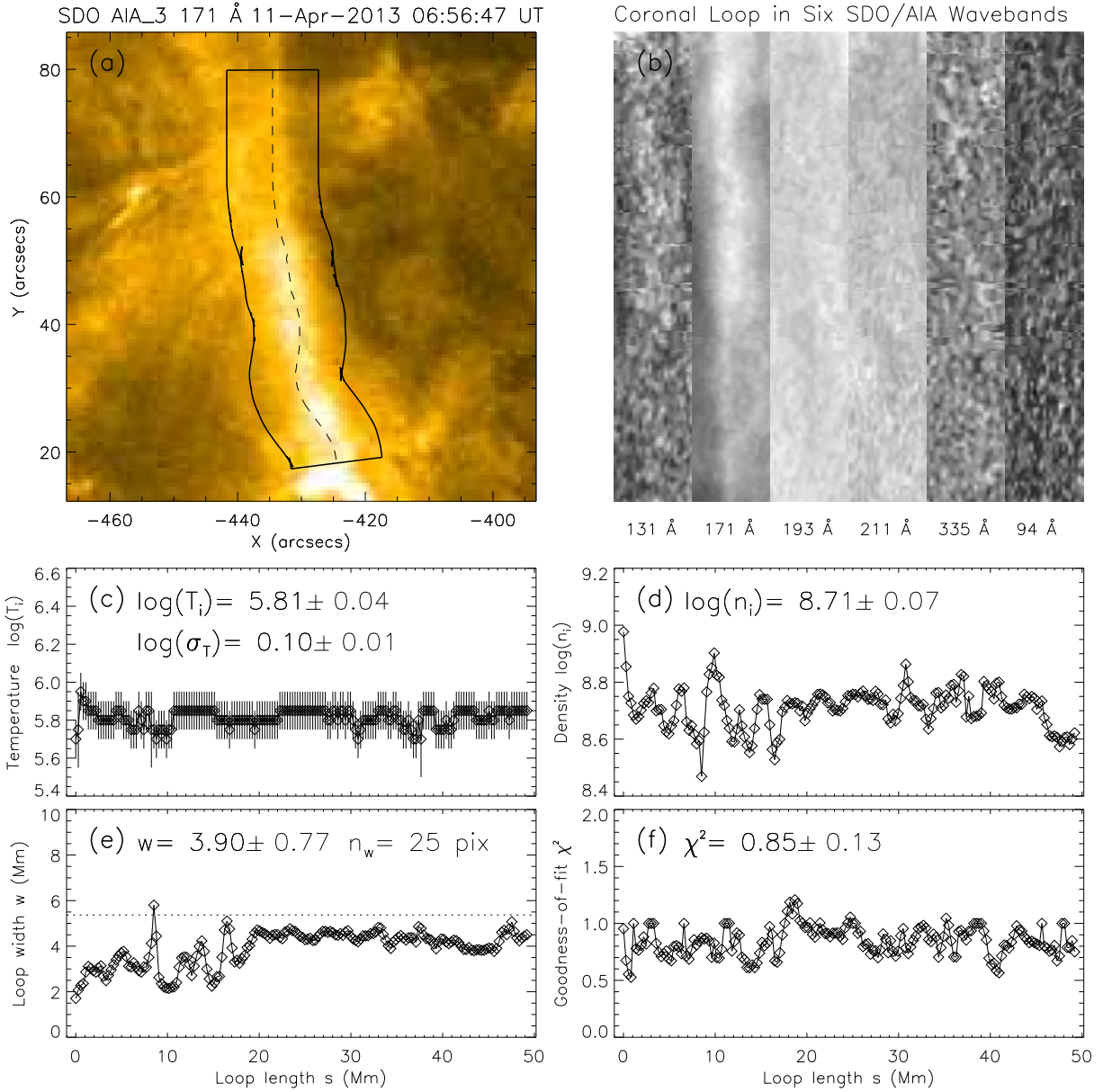}
\caption{(a) Loop path detection using Aschwanden's method. Dashed line represents the detected loop axis. Solid lines surround the detected loop segment with a width of 25 pixels, namely, $15''$. (b) Stretched loop segment in six wavelengths of \textit{SDO}/AIA. (c) Fitted peak temperatures, $T_\mathrm{i}$, and the Gaussian temperature width, $\sigma_T$, along the loop length, $s$. (d) Electron densities, $n_\mathrm{i}$. (e) Loop widths, $w$. The dotted line indicates half of the selected width, $n_w/2 = 12.5$ pixel. (f) Goodness-of-fit, $\chi^2$, for the single Gaussian DEM fitting.
} \label{fig5}
\end{center}
\end{figure}

We adopt the Oriented Coronal CUrved Loop Tracing (OCCULT) code and the single Gaussian forward fitting method proposed by \citet{2013Aschwanden} to detect the loop path and to perform the DEM analysis. Figures~\ref{fig5}(a) and \ref{fig5}(b) show a detected loop segment in the 171~\AA \ waveband and the stretched loop segment in six wavelengths, respectively. The Gaussian loop widths, $\sigma_\mathrm{w}$, are fitted along the cross-sectional profiles in the 171~\AA \ waveband. To substract the background emission, we further fit the cross-sectional profiles in all six \textit{SDO}/AIA wavelengths by a Gaussian (with peak flux $F_\lambda^\mathrm{loop}$ and Gaussian loop width $\sigma_\mathrm{w}$ derived in 171~\AA ) plus a linear background profile. The loop width is estimated as $w = 2 \sqrt{2\ln 2} \sigma_\mathrm{w} \approx 2.35 \sigma_\mathrm{w}$, which is shown in Figure~\ref{fig5}(e). The background-substracted EUV fluxes in six \textit{SDO}/AIA wavelengths, $F_\lambda^\mathrm{loop}$, are used for the single-Gaussian DEM fitting, from which we derive the peak emission measure, $EM_\mathrm{i}$, peak temperature, $T_\mathrm{i}$, and the Gaussian temperature width, $\sigma_T$. The index i denotes that the values are measured inside the coronal loop. The electron density, $n_\mathrm{i}$, is computed as
\begin{eqnarray}
n_\mathrm{i} = \sqrt {\frac{EM_\mathrm{i}}{w}}.
\end{eqnarray}
The distribution of $T_\mathrm{i}$, $\sigma_T$, and $n_\mathrm{i}$ along the loop length is shown in Figures~\ref{fig5}(c) and (d). We find that the average peak temperature is $0.65 \pm 0.06$ MK and the average electron density is $(5.1 \pm 0.8) \times 10^8$ cm$^{-3}$. The average Gaussian temperature width reaches the lower limit of the DEM analysis, which indicates that the loop is almost isothermal. The goodness-of-fit as shown in Figure~\ref{fig5}(f) indicates that the fit results are acceptable.

We note that Figure~\ref{fig5} only shows a subsection of the whole coronal loop. Since the observations of the loop in the EUV wavebands have complicated backgrounds, the OCCULT method of \citet{2013Aschwanden} cannot identify the loop as a whole, but recognize it with several subsections. We have checked all the fitted parameters in other subsections, they are all consistent with those shown in Figure~\ref{fig5}. The DEM analysis does not show a density stratification of the loop, while the observed period ratio ($P_1/P_2 <2.0$) indicates the signature of density stratification along the loop \citep{2007Erdelyi,2008McEwan,2008Verth1}. The departure from hydrostatic equilibrium due to plasma motions in the loop system may attribute to the larger scale height compared to the hydrostatic case \citep{2001Aschwanden,2008Srivastava}. However, present EUV observations do not show clear evidence of plasma motions in the coronal loop system, which is located in a quiescent region with less plasma dynamics as mostly such dynamics occur in active region loops. However, the MHD seismology using the period ratio of kink waves ($P_1/P_2$) clearly demonstrates the signature of density stratification. Therefore,  the detection of a uniform density along the loop using the DEM method may be caused by the improper assumption of the line-of-sight column depth, which is taken as the loop width in this case. Due to the projection effect and non-circular shape of the cross section of the loop, this assumption might be invalid. In conclusion, MHD seismology shows that the period ratio $P_1/P_2$ is below 2.0, which suggests the density stratification in the coronal loop system; however, such stratification is not clearly observed in the DEM analysis using \textit{SDO}/AIA observations due to above mentioned limitations.

\section{Results} \label{sect3}

\subsection{Magnetic Field Strength Derived by the Fundamental Mode}

To compute the magnetic field strength via coronal seismology, we have to determine the ratio between the densities inside and outside the coronal loop. Following Equation~(25) of \citet{2011Aschwanden}:
\begin{eqnarray}
\frac{n_\mathrm{i}}{n_\mathrm{e}} = \frac{1}{2} \left(\frac{L_\mathrm{exc}}{L_\mathrm{osc}} \frac{P_\mathrm{kink}}{T_\mathrm{exc}} \right)^2 - 1 = \frac{1}{2} \left( v_\mathrm{exc} \frac{P_\mathrm{kink}}{L_\mathrm{osc}} \right)^2 - 1,
\end{eqnarray}
we derive that $n_\mathrm{i}/n_\mathrm{e} = 1.3 \pm 0.4$, where $v_\mathrm{exc} = 825.0 \pm 19.5$ km~s$^{-1}$, $P_\mathrm{kink} = 530.2 \pm 13.3$ s, and $L_\mathrm{osc} = 203.8 \pm 13.8$ Mm. The mean magnetic field strength in the coronal loop and its surroundings is given by \citep{1984Roberts,2011Aschwanden}:
\begin{eqnarray}
B = \frac{L_\mathrm{osc}}{P_\mathrm{kink}} \sqrt{8 \pi \mu m_\mathrm{p} n_\mathrm{i} (1 + n_\mathrm{e}/n_\mathrm{i})}, \label{eqn:b}
\end{eqnarray}
where $\mu = 1.2$ is the average molecular weight for coronal abundances \citep{2013Verwichte}, $m_\mathrm{p} = 1.67 \times 10^{-24}$ g is the proton mass. Using the estimated plasma properties under the framework of the fundamental kink mode (Table~\ref{tbl1}), we find that $B = 8.2 \pm 1.0$ G.

Next, we measure the magnetic field strength distribution along the coronal loop using the potential field model. Since the loop has a finite width, we have to carry out some averaging. The magnetic field strength at a normalized position, $s$, is computed as the average strength at the same normalized position of each sample magnetic field line. Their uncertainties are estimated as the standard deviations. We plot the distribution of the total field strength along the coronal loop in Figure~\ref{fig6}(a), which shows that it decreases from the northern foot-point to the southern one monotonically until $s=0.9$. Therefore, the magnetic field is nonuniform and asymmetric along the coronal loop. The average magnetic field strength $B = 8.2 \pm 1.0$~G derived by solar magneto-seismology matches with the values derived by the potential field model. Quantitatively, $B = 8.2 \pm 1.0$ G is between the minimum and maximum field strengths derived by the potential field as shown in Figure~\ref{fig6}(a). Since the magnetic field varies along the coronal loop, \citet{2011Aschwanden} defined an average magnetic field, $\left<B\right> = [\int_0^1 B(s)^{-1} ds]^{-1}$, such that an equivalent loop with the constant magnetic field $\left<B\right>$ would have the same period, $P_1$, as the original loop. Following such a definition, we find that $\left<B\right> \approx 3.6$~G for the coronal loop.

\subsection{Amplitude Profiles of the Spatial Overtones}

Physical parameters along the coronal loop, such as the density and the magnetic field strength, not only affect the period (or frequency) of the loop oscillation, but also the eigenfunction itself, i.e. the amplitude profile. The ratio of the periods provides information about the non-uniformity of the physical parameters along a loop. However, periods of the eigen-modes do not unveil the asymmetry of the physical parameters, while the amplitude profiles provide these information. Therefore, we study the amplitude profiles and estimate the antinode positions in the following study.

\begin{figure}
\begin{center}
\includegraphics[width=0.7\textwidth]{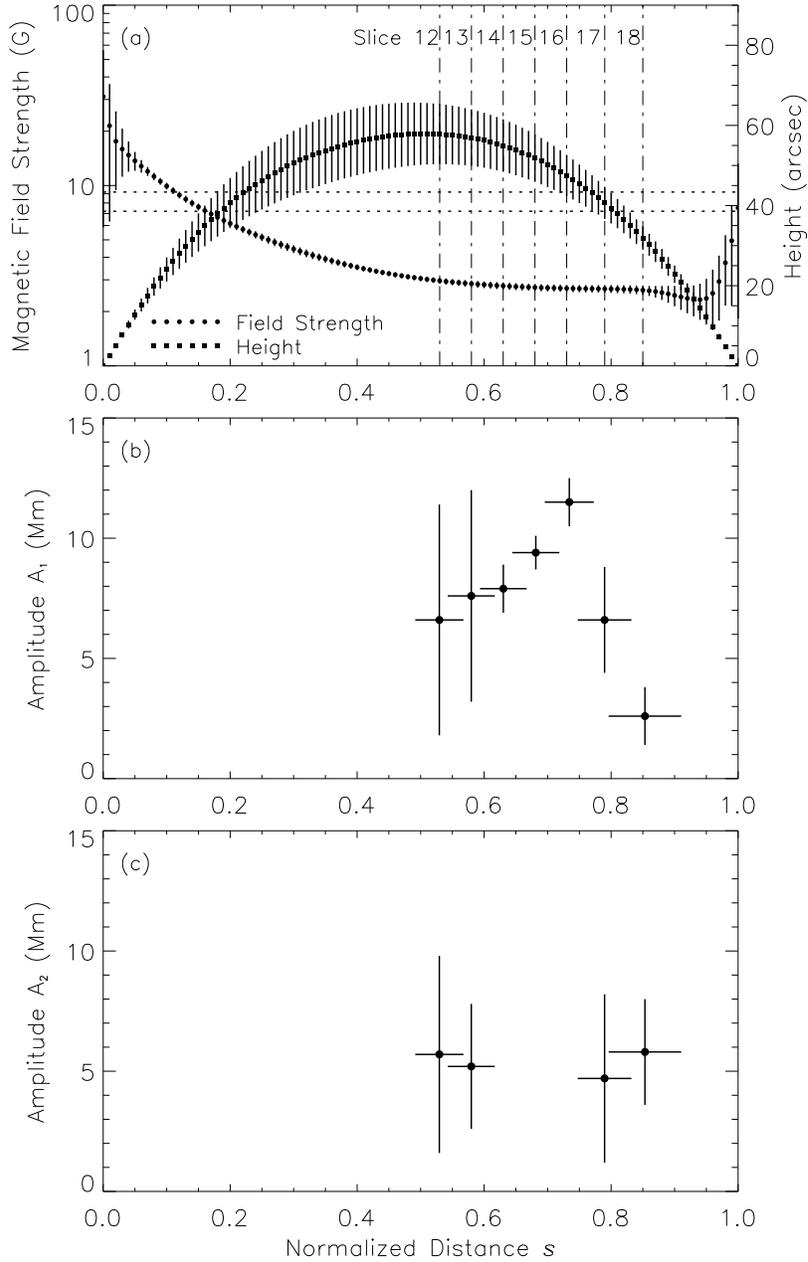}
\caption{(a) Magnetic field strength (filled dot) and loop height (filled square) distributions along the coronal loop. The normalized distance $s = 0$ corresponds to the northern foot-point of the loop. The two dotted lines indicate the range of the magnetic field strength derived by the coronal seismology using the fundamental mode. Dash-dotted lines indicate the positions of the slices. (b) Amplitude distribution of the fundamental mode along the coronal loop. (c) Amplitude distribution of the first overtone along the coronal loop.
} \label{fig6}
\end{center}
\end{figure}

In Section~\ref{sect24}, we have derived the oscillation parameters of the coronal loop for slices 12--18. The first overtones for the northern part and southern part are found along slices 12--13 and 17--18, respectively. The fundamental mode is found along slices 12--18. Next, we have to measure the slice positions in a normalized coordinate system along the coronal loop. We overlay the magnetic loops as shown in Figure~\ref{fig1}(b) on the slices and measure the projection coordinates of the cross points between the slices and the magnetic loops. The projection coordinates can be further converted to the normalized loop coordinate, $s$, since the three-dimensional geometry of the magnetic loops are known from the potential field model. The final position of each slice as shown in Figure~\ref{fig6}(a) is computed as the average of $s$ of all the magnetic loop samples. If the position error is estimated as the standard deviation, it is less than $7\%$ for all the positions of slices 12--18. Figures~\ref{fig6}(b) and (c) show the amplitude distribution of the fundamental mode and the first overtone along the coronal loop, respectively.

To determine the antinode positions, we have to pinpoint the positions where the spatial overtones have the largest amplitude. It is difficult to derive this information from observations because it asks for accurate measurements and fittings of the loop oscillations. Unfortunately, the present measurements have relatively large errors. Here, we provide a rough estimation as follows. First, we find that the amplitudes of the fundamental mode at slices 14--16 are larger than that at other places. The antinode position is estimated as the average positions of the three slices. The uncertainties are taken as the lower bound of the position of slice 14 and the upper bound of the position of slice 16. The estimated antinode position of the fundamental mode is $0.68 \pm 0.09$ (measured along $s$). Next, the first overtones only clearly appear in slices 12--13 in the northern part and slices 17--18 for the southern part. With similar principle for the fundamental mode, the antinode positions for the first overtone are $0.55 \pm 0.06$ for the northern part and $0.82 \pm 0.09$ for the souther part, respectively.

It is known that in a straight and uniform loop, the antinode positions for the first overtone (with two antinodes) and fundamental mode are $s = 0.25$, 0.75, and 0.5, respectively. The shifts of the antinode positions may be caused by the curved and asymmetric loop geometry, the density stratification, and magnetic field strength variation along the coronal loop. We plot the height distribution of the coronal loop in Figure~\ref{fig6}(a), which shows that the loop is almost symmetric. However, the antinode of the fundamental mode is not located at the middle of the loop, different from what is expected for a symmetric coronal loop. Compared to Figure~\ref{fig6}(a), it shifts towards the weak magnetic field region. Moreover, at least for the antinode position of the first overtone in the northern part, it shifts towards the antinode position of the fundamental mode compared to the ideal case for a straight and uniform loop. This clearly indicates the dominance of the magnetic field expansion over the density stratification, which shifts the position of the antinodes of the first overtone towards the loop apex \citep{2007Erdelyi,2009Andries}. We note that the antinode of the fundamental mode is located at the loop apex in the symmetric case of \citet{2007Erdelyi}. However, in asymmetric cases, it is possible that it is not located at the loop apex.

\section{Discussion and Conclusions} \label{sect4}

In the present paper, we study an M6.5 class flare occurred in active region 11719, and the triggered quiescent coronal loop oscillation due to the global fast magneto-acoustic waves associated with this flare on 2013 April 11. We observe the first two modes (fundamental mode and its first overtone) of the horizontal kink waves in the loop. The global fast magneto-acoustic wave propagated with the speed of $v_\mathrm{exc} = 825.0 \pm 19.5$ km~s$^{-1}$, and triggered the MHD oscillations in the observed coronal loop system. Using the automated DEM analyses, we estimate the average electron density and temperature of the loop system as $(5.1 \pm 0.8) \times 10^8$ cm$^{-3}$ and $0.65 \pm 0.06$ MK, respectively. Meanwhile, we use the \textit{SDO}/HMI vector magnetic field to derive the loop geometrical parameters (e.g. the length $203.8 \pm 13.8$ Mm) and field strength along the loop with the potential field model. We find that the magnetic field derived by the principle of the solar magneto-seismology ($B = 8.2 \pm 1.0$ G) using the oscillatory properties of the fundamental kink mode, precise loop geometry, and plasma parameters, matches with that derived by the potential field extrapolation using the \textit{SDO}/HMI vector magnetic field.

However, the magnetic field strength derived by the magneto-seismology, $B = 8.2 \pm 1.0$ G, does not equal to the average strength along the coronal loop, $\left<B\right> \approx 3.6$~G, following the definition in \citet{2011Aschwanden}, where the authors found that the former is less than the latter. \citet{2013Verwichte} further proposed a weighted average of the magnetic field strength to improve their results. This method would decrease the average magnetic field since it gives larger weights for the weak field close to the loop apex than for the strong field close to the loop footpoints. In our case, this weighted average would enlarge the discrepancy. We think that one source of the discrepancy comes from the measurements of all the geometrical and physical parameters and the magnetic field model itself. Another source might arise from the computation of the average magnetic field, which neglects the variation of the density along the coronal loop. The density stratification would increase the weight for the strong magnetic field close to the footpoints.

The seismological field strength obtained only with the fundamental mode is a mean field over the whole coronal loop. However, under the realistic conditions in the solar atmosphere when both the density and magnetic field varies along the coronal loop, both the frequency and spatial properties of the loop oscillations are different from that of uniform models. On the one hand, the density along the coronal loop should decrease to its top. Although the DEM analysis does not show clear density stratification in this case, it is most probably caused by the assumption of the line-of-sight column depth being approximately the loop width. This assumption may not be valid due to the projection effect and the non-circular shape of the cross section of the loop. The potential field model shows that the magnetic field strength along the coronal loop is nonuniform and asymmetric. On the other hand, the period ratio of the fundamental mode and the first overtone is less than the canonical value of 2, i.e. $P_1/P_2 < 2.0$. The antinode positions of both the fundamental mode and the first overtone shift towards the weak field region. Especially, the antinode of the first overtone in the northern part shifts towards the antinode of the fundamental mode compared to ideal cases where the loop is straight and uniform.

Some aspects of the previous findings can be explained qualitatively by existing theories of solar magneto-seismology. It has been found that the period ratio, $P_1/P_2$, can be affected by various factors, such as the finite tube width and curvature \citep{2006McEwan,2004VanDoorsselaere}, density stratification \citep{2005Andries,2005Dymova,2006Dymova,2008Verth1}, magnetic field variation along the flux tube \citep{2008Verth1,2008Ruderman}, and the temperature difference between plasma inside and outside the loop \citep{2012Orza}. In our case, the ratio between the radius and the length of loop, $0.5w/L_\mathrm{osc}$, is about 0.01; therefore, the effect of the finite tube width and curvature can be safely neglected. The temperature of the coronal loop ($0.65 \pm 0.06$ MK) is lower than that of its ambient environment, which is $\sim 1$ MK measured from the DEM analysis. Following the formulae in \citet{2012Orza}, we find that the temperature difference effect decreases the period ratio, $P_1/P_2$, by about 20\% referred to the case with uniform temperature. Meanwhile, both the density stratification effect and the magnetic field variation along the loop affect the period ratio. Solar magneto-seismology theories show that $P_1/P_2$ decreases as the density stratification effect increases (equivalently, the density scale height decreases), while $P_1/P_2$ increases as the magnetic field decreases along the height \citep{2008Verth}. Therefore, the magnetic field variation along the height usually has the opposite effect as the density stratification and the temperature difference do on the period ratio \citep{2009Andries}. In our case, since $P_1/P_2 < 2.0$, the density stratification and the temperature difference effects are larger than the magnetic field variation effect on the period ratio.

Physical parameters along a loop determine not only the periods of the oscillations, but also their eigenfunctions. \citet{2007Erdelyi} suggested that the information contained in the eigenfunction can be used in the solar magneto-seismology. Theories based on such an idea in the spatial domain have been developed further by \citet{2007Verth} and \citet{2009Andries}. \citet{2007Verth} and \citet{2008Verth} find that the antinodes of the first overtone shift towards the foot-points with the density stratification effect, but towards the loop apex with the magnetic field expansion effect \citep[see also][]{2009Andries}. Our observations indicate that the antinode of the first overtone in the northern part shift towards the antinode of the fundamental mode; therefore, the magnetic field expansion effect dominates the density stratification effect in this case for determining the antinode position. Compared to the result for the periods, antinode position is more sensitive to the magnetic field distribution than the density stratification, while period ratio is more sensitive to the density stratification and temperature difference than the magnetic field distribution in this case.

Although part of our findings can be explained qualitatively by existing theories of solar magneto-seismology, none of the theories can be applied to the observations quantitatively to infer the physical parameters, such as the density scale height and magnetic field expansion. The reason is that almost all the available theories have been developed with the assumption that a coronal loop is symmetric in both geometry and thermal property. The only exception is \citet{2013Orza}, where the geometrical asymmetry is considered. To develop a practical spatial MHD seismology, models with more specific geometries and physical parameter settings need to be developed. For example, one needs a theory to determine the antinode positions of the fundamental mode and the overtones for a coronal loop in real conditions such as asymmetric magnetic field and density distributions along the loop and temperature difference between the coronal loop and its surroundings.

In conclusion, this study shows that the combined observations of eigen-frequencies and eigenfunctions of kink wave overtones put more constraints on coronal seismology, especially when both the density and magnetic field strengths vary along the coronal loops and the temperatures are different inside and outside them. Magnetic extrapolation is also a useful tool to determine the loop geometry and the magnetic field distribution. Our present observations clearly match the basic theoretical scenario of the multiple overtones of the kink waves excited in the coronal loops. To fully take the advantage of solar magneto-seismology, both MHD wave observation and wave theory need to be advanced. From the observational side, future studies should focus on further reducing the observational errors. From the theoretical side, it is necessary to consider how more general physical parameters along the coronal loop affect both the frequency and spatial properties of the loop oscillations.

\acknowledgments

The authors thank the anonymous referee for constructive suggestions that improve the quality of the paper. YG Thanks Dr. Ballai for useful discussions. YG, QH, XC, PFC, and MDD are supported by NSFC (11203014, 10933003, and 11025314) and NKBRSF (2011CB811402 and 2014CB744203). RE is thankful to the NSF of Hungary (OTKA, Ref. No. K83133) and acknowledges M. K\'eray for patient encouragements. AKS and RE acknowledge the Royal Society International Exchange Scheme grant (2014-2016) to carry out the joint research.  AKS thanks PFC for his invitation to Nanjing University and Shobhna for her patient encouragements.




\clearpage

\begin{table}
\caption{Displacement and its change rate, reference time, initial amplitude, period, initial phase, and damping time at different positions and wave modes.} \label{tbl1}
\begin{tabular}{l l l l} \\
\hline \hline
Parameters              & Slice 12           & Slice 14           & Slice 17         \\
\hline
$A_{00}$ (Mm)           & $31.0 \pm 0.7$     & $30.2 \pm 0.3$     & $32.3 \pm 0.3$   \\
$A_{01}$ (km s$^{-1}$)  & $-4.7 \pm 1.1$     & $-1.8 \pm 0.1$     & $0 ^\mathrm{a}$  \\
$t_0$ (UT)              & 07:08:46           & 07:08:48           & 07:08:48         \\
$A_1$ (Mm)              & $6.6 \pm 4.8$      & $7.9 \pm 1.0$      & $6.6 \pm 2.2$    \\
$P_1$ (s)               & $520 ^\mathrm{a}$  & $530.2 \pm 13.3$   & $519.9 \pm 55.3$ \\
$\phi_{01}$ ($^\circ$)  & $21.9 \pm 30.5$    & $31.7 \pm 9.2$    & $73.5 \pm 33.5$  \\
$\tau_1$ (s)            & $203.7 \pm 131.1$  & $657.8 \pm 107.4$   & $621.4 \pm 307.1$\\
$A_2$ (Mm)              & $5.7 \pm 4.1$      & -                  & $4.7 \pm 3.5$    \\
$P_2$ (s)               & $300.4 \pm 27.7$   & -                  & $334.7 \pm 22.1$ \\
$\phi_{02}$ ($^\circ$)  & $237.7 \pm 36.4$   & -                  & $28.9 \pm 27.0$  \\
$\tau_2$ (s)            & $329.5 \pm 180.8$  & -                  & $611.9 \pm 540.9$\\
\hline
\multicolumn{4}{l}{$^\mathrm{a}$ These values are prescribed. See text for more details.}
\end{tabular}
\end{table}
\end{document}